\documentclass[a4paper,12pt]{article}
\usepackage{graphicx}
\usepackage{setspace}
\usepackage{caption}
\usepackage{subfigure}
\usepackage{url}
\usepackage{natbib}
\usepackage{multirow}
\usepackage{color}

\begin{document}

\title{featureCounts: an efficient general-purpose program for assigning sequence reads to genomic features}
\author{Yang Liao\,$^{1,2}$, Gordon K Smyth\,$^{1,3}$ and Wei Shi\,$^{1,2}$\footnote{To whom correspondence should be addressed. Email: shi@wehi.edu.au}}
\date{}
\maketitle
{\noindent$^{1}$Bioinformatics Division, The Walter and Eliza Hall Institute of Medical Research, 1G Royal Parade, Parkville, VIC 3052,
$^{2}$Department of Computing and Information Systems,
$^3$Department of Mathematics and Statistics, The University of Melbourne, Parkville, VIC 3010, Australia
}

\begin{center}
\Large\bf{Abstract}
\end{center}

Next-generation sequencing technologies generate millions of short sequence reads, which are usually aligned to a reference genome.
In many applications, the key information required for downstream analysis is the number of reads mapping to each genomic feature, for example to each exon or each gene.
The process of counting reads is called read summarization.
Read summarization is required for a great variety of genomic analyses but has so far received relatively little attention in the literature.
We present \emph{featureCounts}, a read summarization program suitable for counting reads generated from either RNA or genomic DNA sequencing experiments.
\emph{featureCounts} implements highly efficient chromosome hashing and feature blocking techniques.
It is considerably faster than existing methods (by an order of magnitude for gene-level summarization) and requires far less computer memory.
It works with either single or paired-end reads and provides a wide range of options appropriate for different sequencing applications.
\emph{featureCounts} is available under GNU General Public License as part of the Subread (\url{http://subread.sourceforge.net}) or Rsubread (\url{http://www.bioconductor.org}) software packages.

\section{Introduction}

Next-generation (next-gen) sequencing technologies are revolution-izing biology by providing the ability to sequence DNA at unprecendented speed \citep{schuster2007nextgen,metzker2009sequencing}.
The computational problem of mapping short sequence reads to a reference genome has received enormous attention in the past few years \citep{li2009bwa,langmead2009bowtie,fonseca2012mappingtools,marco2012gem,liao2013subread}, and the rapid development of fast and reliable aligners is one of the success stories of bioinformatics.
Raw aligner output however is not usually sufficient for biological interpretation.
Read mapping results have to be summarized in terms of read coverage for genomic features of interest before they can be interpreted biologically.
One of the most ubiquitious operations that forms part of many next-gen analysis pipelines is to count the number of reads overlapping pre-determined genomic features of interest.
Depending on the next-gen application, the genomic features might be exons, genes, promotor regions, gene bodies or other genomic intervals.
Read counts are required for a wide range of count-based statistical methods for differential expression or differential binding analysis \citep{oshlack2010from}.

Despite its importance in genomic research, the read counting problem has received little specific attention in the literature.
The problem may appear superficially simple but in practice has many subtleties.
Read count programs need to accommodate both DNA and RNA sequencing as well as single and paired-end reads.
The reads or paired-end fragments to be counted may incorporate insertions, deletions or fusions relative to the reference genome, and these complications need to be accounted for when comparing the location of each read or fragment to each possible target genomic feature.
When the number of features is large, the computational cost of read counting can be comparable to that of the read alignment step.

DNA sequence reads arise from a variety of technologies including ChIP-seq for transcription factor binding sites \citep{valouev2008tfchipseq}, ChIP-seq for histone marks \citep{park2009chipseq}, and assays that detect DNA methylation \citep{harris2010methylation}.
The genomic features of interest for DNA reads can usually be specified in terms of simple genomic intervals.
For example, \cite{pal2013ezh2} counted reads associated with histone marks by gene promotor regions and by whole gene bodies.
\cite{ross2012differentialbinding} counted reads overlapping with intervals identified by a peak caller \citep{zhang2008macs}.

Counting RNA-seq reads is somewhat more complex because of the need to accommodate exon splicing.
One way is to count reads overlapping each annotated exon, an approach that can be used to test for alternative splicing between experimental conditions \citep{anders2012dexseq,reyes2013drift}.
Another common approach is to summarize counts at the gene-level, by counting all reads that overlap any exon for each gene \citep{anders2013count,bhattacharyya2013genome,man2013irf4}.
Gene annotation from RefSeq \citep{pruitt2012refseq} or Ensembl \citep{flicek2012ensembl} is often used for this purpose.

Read counts provide an overall summmary of the coverage for the genomic feature of interest.
In particular, gene-level counts from RNA-seq provide an overall summary of the expression level of the gene but do not distinguish between isoforms when multiple transcripts are being expressed from the same gene.
Reads can generally be assigned to genes with good confidence, but estimating the expression levels of individual isoforms is intrinsically more difficult because different isoforms of the gene typically have a high proportion of genomic overlap.
A number of model-based methods have been developed that attempt to deconvolve the expression levels of individual transcripts for each gene from RNA-seq data, essentially by leveraging information from reads unambiguously assigned to regions where isoforms differ \citep{trapnell2010cufflinks,li2011rsem}.
This article concentrates on the read count problem, which is generally applicable even when the sequencing depth is not sufficient to make transcript level analysis reliable.
Many statistical analysis methods have been developed to detect differentially expression or differential binding on the basis of read counts \citep{mccarthy2012edgerglm,anders2010deseq,li2012poissonseq,hardcastle2010bayseq,auer2011tspm,wu2013dss}.
Recent comparisons have concluded that the read count methods perform well relative to model-based methods for the purposes of gene level differential expression \citep{nookaew2012comparison,rapaport2013comprehensive} or detection of splice variation \citep{anders2012dexseq}.

Only a handful of general purpose read count software tools are currently available.
The software packages \emph{GenomicRanges} \citep{genomicranges} and \emph{IRanges} \citep{iranges}, developed by the core team of the Bioconductor project \citep{gentleman2004bioconductor}, include functions for counting reads that overlap genomic features.
The \emph{countOverlaps} function of \emph{IRanges} is designed for counting reads overlapping exons or other simple genomic regions,
while the \emph{summarizeOverlaps} function of \emph{GenomicRanges} is designed for counting reads at the gene level.
Another tool is the \emph{htseq-count} script distributed with the HT-Seq Python framework for processing RNA-seq or DNA-seq data \citep{htseq}.
All of these are popular and well-tested software tools, but all make extensive use of programming in the interpreted computer languages R or Python and none are fully optimized for efficiency and speed.
\emph{BEDTools} is a popular tool for finding overlaps between genomic features that can be used to count overlaps between reads and features \citep{quinlan2010bedtools}.
It is fully implemented in the compiled language C++, making it faster than the above tools.
It is however not specifically designed for RNA-seq data, so can count reads for exons or interval features only, similar to \emph{countOverlaps}.

This article presents a highly optimized read count program called \emph{featureCounts}.
\emph{featureCounts} can be used to quantify reads generated from either RNA or DNA sequencing technologies in terms of any type of genomic feature.
It implements chromosome hashing, feature blocking and other strategies to assign reads to features with very high efficiency.
It supports multithreading, which provides further speed improvements on large data problems.
It is available either as a Unix command or as a function in the R package \emph{Rsubread}.
In either case, all the core functionality is written in the C programming language.
The R function is a wrapper for the compiled C code that provides the convenience of the R programming environment without sacrificing any of the efficiency of the C implementation.

\section{Data formats and inputs}

\subsection{Input data}

The data input to \emph{featureCounts} consists of (i) one or more files of aligned reads in either SAM or BAM format \citep{li2009samtools} and (ii) a list of genomic features in either General Feature Format (GFF) \citep{gff} or Simplified Annotation Format (SAF) \citep{saf}.
The read input format (SAM or BAM) is automatically detected and so does not need to be specified by the user.
Both the read alignment and the feature annotation should correspond to the same reference genome, which is a set of reference sequences representing chromosomes or contigs.
For each read, the SAM or BAM file gives the name of the reference chromosome or contig to which the read mapped, the start position of the read on the chromosome or contig, and the so-called CIGAR string giving the detailed alignment information including insertions and deletions and so on relative to the start position.

The genomic features can be specified in either GFF or SAF format.
The SAF format is the simpler and includes only five required columns for each feature: feature identifier, chromosome name, start position, end position and strand.
These five columns provide the minimal sufficient information for read quantification purposes.
In either format, the feature identifiers are assumed to be unique, in accordance with commonly used Gene Transfer Format (GTF) refinement of GFF \citep{gtf}.

The number of reference sequences may be small or large depending on the application.
For well established genomes, the number of reference sequences is equal or very close to the number of chromosomes.
The number of reference sequences can be however much larger for genomes with incomplete or low quality assemblies because each contig becomes a reference sequence.
RNA-seq reads are sometimes aligned to the transcriptome instead of to the genome.
In this case there may be hundreds of thousands of transcripts and each transcript become a reference sequence.

\emph{featureCounts} supports strand-specific read counting if strand-specific information is provided.
Read mapping results usually include mapping quality scores for mapped reads.
Users can optionally specify a minimum mapping quality score that the assigned reads must satisfy.

\subsection{Single and paired-end reads}

Reads may be paired or unpaired.
If paired reads are used, then each pair of reads defines a DNA or RNA fragment bookended by the two reads.
In this case, \emph{featureCounts} will count fragments rather than reads.
By default, paired reads are assumed to be in consecutive positions in the SAM or BAM file.
Otherwise, \emph{featureCounts} can optionally sort reads by name to ensure this is the case.

\subsection{Features and meta-features}

Each feature is an interval (range of positions) on one of the reference sequences.
We also define a meta-feature to be a set of features representing a biological construct of interest.
For example, features often correspond to exons and meta-features to genes.
Features sharing the same feature identifier in the GFF or SAF annotation are taken to belong to the same meta-feature.
\emph{featureCounts} can summarize reads at either the feature or meta-feature levels.

\begin{figure*}[t]
\centering
\includegraphics[scale=0.6]{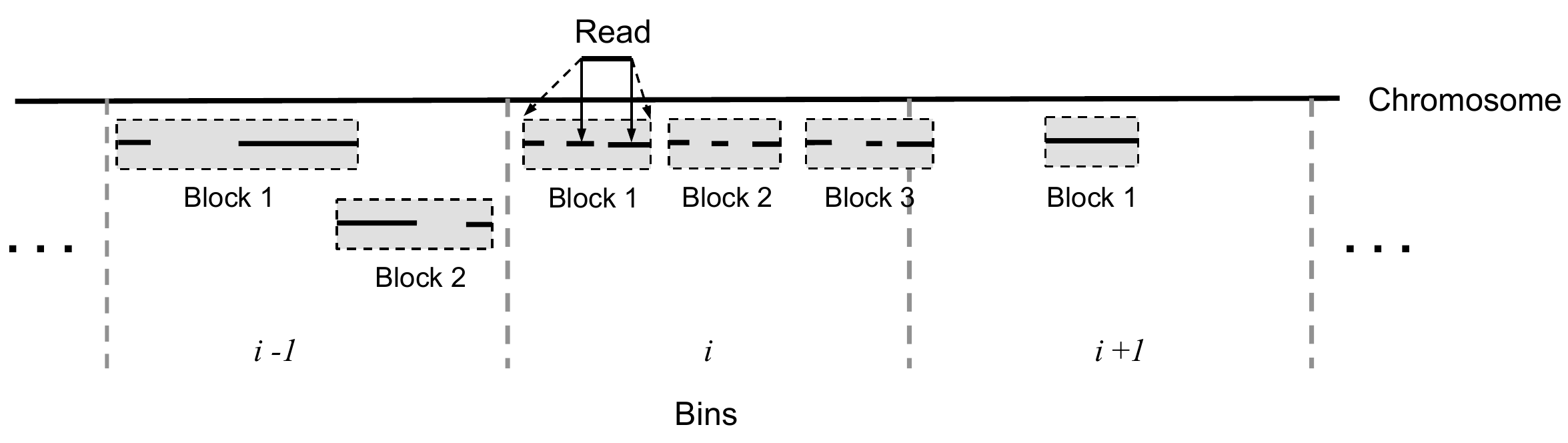}
\caption{Genomic bins and feature blocks.
Each chromosome is divided into 128kb bins.
Features (solid lines) are assigned to bins according to their start positions and grouped into blocks (grey boxes) within each bin.
Query reads are compared with genomic bins, then with blocks (dashed arrows) and finally with features (solid arrows).
The query read in the figure overlaps with two features in the first block of bin $i$.}
\label{fig:index}
\end{figure*}

\section{Algorithm}

\subsection{Overlap of reads with features}

\emph{featureCounts} preforms precise read assignment by comparing mapping location of every base in the read or fragment with the genomic region spanned by each feature.
It takes account of any gaps (insertions, deletions, exon-exon junctions or fusions) that are found in the read.
It calls a hit if any overlap (1bp or more) is found between the read or fragment and a feature.

A hit is called for a meta-feature if the read or fragment overlaps any component feature of the meta-feature.

\subsection{Multiple overlaps}

A multi-overlap read or fragment is one that overlaps more than one feature, or more than one meta-feature when summarizing at the meta-feature level.
\emph{featureCounts} provides users with the option to either exclude multi-overlap reads or to count them for each feature that is overlapped.
The decision whether or not to counting these reads is often determined by the experiment type.
We recommend that reads or fragments overlapping more than one gene are not counted for RNA-seq experiments, because any single fragment must originate from only one of the target genes but the identity of the true target gene cannot be confidently determined.
On the other hand, we recommend that multi-overlap reads or fragments are counted for most ChIP-seq experiments because epigenetic modifications inferred from these reads may regulate the biological functions of all their overlapping genes \citep{pal2013ezh2}.

Note that, when counting at the meta-feature level, reads that overlap multiple features of the same meta-feature are always counted exactly once for that meta-feature, provided there is no overlap with any other meta-feature.
For example, an exon-spanning read will be counted only once for the corresponding gene even if it overlaps with more than one exon.

\subsection{Chromosome hashing}

The first step of the \emph{featureCounts} algorithm is to generate a hash table for the reference sequence names.
This allows the reference sequence names found in the SAM files and GFF annotation to be matched very quickly.
This is particularly useful when there is a large number of reference sequences.
After matching reads and features by reference sequence, subsequent analysis can proceed for each reference sequence separately.

\subsection{Genome bins and feature blocks}

After hashing the reference sequence names, the features in each reference sequence are sorted by their start positions (leftmost base positions).
A two-level hierarchy is then created for each reference sequence.
First, the reference sequence is divided into non-overlapping 128kb bins and features are assigned to bins according to their start positions.
Within each bin, equal numbers of consecutive features are grouped into blocks (Fig.~\ref{fig:index}).
The number of blocks in a bin is the square-root of number of features in that bin (rounded up to the next whole number).
This ensures that the number of features in a block is nearly equal to the number of blocks in a bin, an optimal setting for a hierarchical search.

The use of a hierarchical data structure (features within blocks within bins) is a key component of the \emph{featureCounts} algorithm.
It facilitates rapid read assignment by quickly narrowing down the genomic region that could contain features overlapping with the query read.
The query read is compared first with genomics bins, then with feature blocks within any overlapping bins, then with features in any overlapping blocks.
Instead of using multiple levels of bins \citep{Kent2002,quinlan2010bedtools}, the algorithm uses only one level of bins in combination with the feature blocks.
Finally, the algorithm decides how to assign the read according to which level of summarization is being performed (feature level or meta-feature level) and whether the read is allowed to overlap with more than one target at that level.

\section{Implementation}

The \texttt{featureCounts} command in the \emph{Subread} package for Unix is written entirely in the C programming language.
The memory footprint is minimized by holding in memory only the feature annotation data required at each stage of the computation.
The C code supports multithreading, and the user can specify the number of threads to be used.
One thread is the default.

The R function \texttt{featureCounts} in the \emph{Rsubread} package for R is a wrapper for the same compiled C code as for the Unix command line.
The R function provides the convenience of the R programming environment without sacrificing any of the efficiency of the C implementation.
It produces a data object in R that can be input directly into R-based statistical analysis software such as \emph{edgeR} \citep{robinson2010edgeR} or \emph{limma} \citep{law2013voom} that are designed to analyse next-gen read counts.

\section{Performance on RNA-seq data}

\subsection{Data and annotation}

First we compare the performance of \emph{featureCounts} with existing software tools for counting RNA-seq reads at the gene level.
As an example case study, we use RNA-seq data that was generated as part of the SEQC (SEquencing Quality Control) project, the third stage of the MAQC project \citep{shi2006maqc}.
This data consists of 6.8 million pairs of 101bp reads generated by sequencing a sample of Universal Human Reference RNA on an Illumina HiSeq 2000.

The SEQC RNA-seq dataset was aligned to the human genome GRCh37 using the \emph{Subjunc} aligner included in the \emph{Subread} package \citep{liao2013subread,subread_sf,Rsubread}.
We used \emph{Subjunc} for this analysis because it explicitly identifies exon-exon junctions and outputs the mapping location of every base of every read including those that span multiple exons.
This allowed us to examine rigorously the ability of the read count programs to count reads spanning multiple exons as well as reads falling within exons.

Genes and exons were defined as in the NCBI human RefSeq annotation build 37.2.
This included 25,702 genes and 225,071 exons.

Counts were summarized at the gene level.
This is, exons were defined to be features, genes were defined as meta-features, and quantification was at the meta-feature level.
As this is RNA-seq data, reads or fragments that overlapped multiple genes should be excluded from the counts.

\subsection{Comparative performance when counting reads}

To demonstrate \emph{featureCounts} on single-end reads, the first evaluation uses only the first read from each read pair.
Table~\ref{tab:rnaseq} compares the performance of \emph{featureCounts} to that of the \emph{summarizeOverlaps} function of the \emph{GenomicRanges} package and to the \emph{htseq-count} script.
\emph{featureCounts} and \emph{summarizeOverlaps} yielded identical counts for every gene (Table~\ref{tab:rnaseq}, column 2).

\begin{figure}[h]
\centering
\subfigure[Single-end]{
\label{fig:vennSE}
\includegraphics[width=0.5\linewidth]{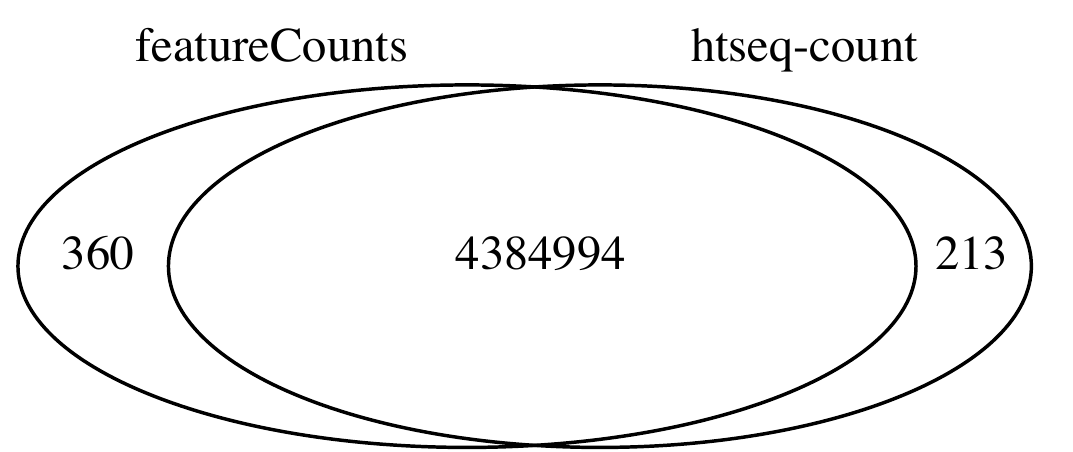}
}
\vspace{0.1cm}
\subfigure[Paired-end]{
\label{fig:vennPE}
\includegraphics[width=0.5\linewidth]{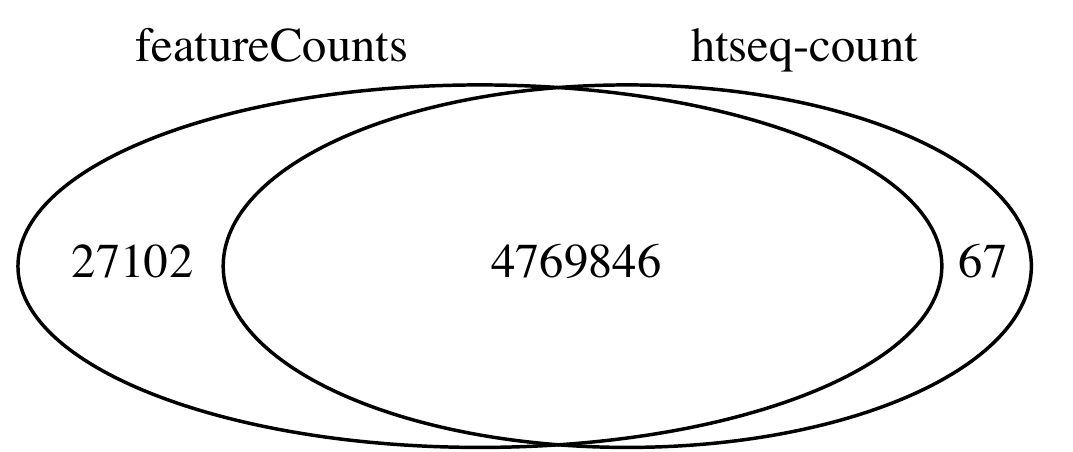}
}
\caption{Concordance between \emph{featureCounts} and \emph{htseq-count} regarding assignment of reads (a) or fragments (b) to genes.
The dataset is the same as for Table~\ref{tab:rnaseq}.
The Venn diagram overlap gives the number of reads or fragments assigned by both methods to the same gene.
The remaining counts give the number of reads or fragments assigned by one method to some genes but not by the other method.}
\label{fig:venn}
\end{figure}

\emph{htseq-count} counted slightly fewer reads than \emph{featureCounts} and \emph{summarizeOverlaps}.
We had a close look at the summarization results for each read given by \emph{htseq-count} and \emph{featureCounts} and found that only a small number of reads were assigned to different genes by the two methods ({Fig.~\ref{fig:vennSE}}).
By comparing the regions these reads were mapped to with the features regions, we identified the reason causing this discrepany.
\emph{htseq-count} takes the right-most base position of each feature as an open position and excludes it from read summarization, whereas \emph{featureCounts} and \emph{summarizeOverlaps} take it as a closed position and includes it in their summarizations.
The GFF specification states that the start and end positions of features are inclusive \citep{gff}, so the interpretation of \emph{featureCounts} and \emph{summarizeOverlaps} appears to be correct.
GFF is the only annotation format supported by \emph{htseq-count}.
We modified the annotation file provided to \emph{htseq-count} by adding one to the right-most position of each exon to let \emph{htseq-count} include these positions.
After this modification, \emph{htseq-count} yielded identical counts to \emph{featureCounts} and \emph{summarizeOverlaps}.

Here and all subsequent comparisons, the software tools were tested on a HP Blade supercomputer with 64 AMD Opteron 2.3GHz CPUs and 512 GB of memory.
All programs were run using a single CPU without multithreading.
Comparisons used software packages \emph{Subread} 1.4.1, \emph{Rsubread} 1.11.15, \emph{GenomicRanges} 1.12.5, \emph{IRanges} 1.18.4, \emph{htseq-count} 0.5.4p3 and \emph{BEDTools} 2.17.0.

\begin{table}[t]
\caption{Performance results on the SEQC RNA-seq data.
Results are given for genewise counts of either single-end reads or paired-end fragments.
\emph{featureCounts} yields the same read counts as \emph{summarizeOverlaps} but is much faster and memory efficient.
\emph{summarizeOverlaps} counts fewer fragments because it excludes read-pairs with only one end successfully mapped.
\emph{htseq-count} counts slightly fewer reads or fragments than \emph{featureCounts} because it interprets GFF annotation differently and calls more ambiguously assigned fragments.
\label{tab:rnaseq}}
{\begin{tabular}{lp{2cm}p{2.5cm}p{2.2cm}p{2.2cm}}
\hline
Method & \# reads & \# fragments & Time & Memory \\
&&& (Mins) & (MB) \\
\hline
\emph{featureCounts}			& 4,385,354 & 4,796,948 & 1.0 & 16  \\
\rule{0pt}{3ex}\emph{summarizeOverlaps} & 4,385,354 & 3,942,439 & 12.1 & 3,400\\
(whole genome at once) \\
\rule{0pt}{2.5ex}\emph{summarizeOverlaps} & 4,385,354 & 3,942,439 & 41.7 & 661 \\
(by chromosome) \\
\rule{0pt}{3ex}\emph{htseq-count}       & 4,385,207 & 4,769,913 & 22.7 & 101 \\
\hline\\
\end{tabular}}
{\linebreak
The table gives the total number of reads counted when using single-end reads and the total number of fragments counted when using paired-end reads.
Running time and memory usage are for fragment summarization.
\emph{featureCounts} was set to exclude reads or fragments overlapping multiple genes.
\emph{summarizeOverlaps} and \emph{htseq-count} were run in `union' mode.
Results are shown for countOverlaps (i) when run on the whole genome at once and (ii) when run chromosome by chromosome.}
\end{table}

\subsection{Comparative performance when counting fragments}

We went on to compare the same methods for counting paired-end fragments, using the full SEQC paired-end data.
\emph{summarizeOverlaps} counted far fewer fragments than \emph{featureCounts} and \emph{htseq-count} (Table~\ref{tab:rnaseq}, column 3).
The main reason for this discrepancy is that \emph{summarizeOverlaps} requires fragments to have both end reads successfully mapped before assigning them to genes, whereas \emph{featureCounts} and \emph{htseq-count} do not have such a requirement, i.e., they can assign fragments just one end mapped.
With a read length of 101bp, fragments with only one end mapped can still have relatively high mapping confidence.
Counting such fragments seems likely to benefit downstream analyses.
Many aligners report fragments that have only one end mapped, including Subread, Subjunc \citep{liao2013subread}, Bowtie \citep{langmead2009bowtie} and TopHat \citep{trapnell2009tophat}.
Almost all (92\%) of fragments counted by \emph{featureCounts} but not by \emph{summarizeOverlaps} were assigned to genes that also had at least 100 assigned fragments with both ends mapped.
This shows that the fragments were assigned to genuinely expressed genes, giving confidence that the extra fragments have been assigned correctly.
Only 0.1\% of extra fragments counts by \emph{featureCounts} were assigned to genes not supported by any fragment with both ends mapped.

\emph{htseq-count} also counts fewer fragments than \emph{featureCounts} in this evaluation (Fig.~\ref{fig:vennPE}).
Running \emph{htseq-count} in `IntersectionStrict' or `IntersectionNotEmpty' modes instead of `Union' mode did not cause it count more fragments.

\emph{featureCounts} can distinguish those features that overlap with different numbers of reads from the same fragment.
For example, if two genes were found to both overlap with a fragment but one gene was found to overlap with only one read and the other with both reads from that fragment, \emph{featureCounts} will assign that fragment to the gene overlapping with both reads.
However, \emph{htseq-count} will take this fragment as ambiguous and will not assign it to any gene.
This is the main reason why \emph{featureCounts} counted slightly more fragments than \emph{htseq-count}.
\emph{featureCounts} uses the size of overlap (in terms of reads) to recover those `ambiguous' fragments.
For this dataset, more than 86\% of fragments assigned by \emph{featureCounts} but not by \emph{htseq-count} were assigned to genes that already had at least 100 unambiguous fragments assigned by both methods.
Only 0.2\% of extra fragments assigned by \emph{featureCounts} were not supported by commonly assigned fragments.
This again shows that the extra fragments are being assigned to genuinely expressed genes, suggesting that the extra fragments are likely to have been correctly assigned.


Table~\ref{tab:rnaseq} (columns 4 and 5) shows that \emph{featureCounts} was considerably faster ($>$10-folds) and more memory efficient than the other programs.
\emph{summarizeOverlaps} was also run chromosome by chromosome to save memory.
That is, reads were split into groups according to the chromosomes they were mapped to and each group of reads was summarized separately.
But it still used 20 times as much memory as \emph{featureCounts}.

\section{Performance on ChIP-seq data}

\subsection{Data and annotation}

Now we compare the performance of \emph{featureCounts} with existing software tools for counting gDNA-seq reads at the feature level.
As an example case study, we use a ChIP-seq dataset that was generated as part of a study of global changes in the mammary stem cell epigenome under hormone perturbation \citep{pal2013ezh2}.
Specifically the dataset was generated to find genomic regions associated with the H3K27me3 epigenetic histone mark (tri-methylation of the histone H3 lysine 27) in mouse mammary stem cells.
This dataset consists of 15 million pairs of 35bp DNA reads generated by an Illumina Genome Analyzer IIx.
The study analysed the total number of fragments mapped to the broad region of each gene, where the broad region is defined to be the entire gene body from first to last base plus the 3kb region immediately upstream from the transcription start of the gene representing the putative promotor region \citep{pal2013ezh2}.

The read mapping and annotation used here follows the original study.
Reads were mapped to the mouse genome (mm9) using the \emph{Subread} aligner \citep{liao2013subread}.
Fragments were included in the evaluation only if both paired-reads were successfully mapped to the genome
and if the fragment defined by the end reads was between 50bp and 500bp long.
The transcription start and end positions for each gene were obtained from the NCBI mouse RefSeq annotation (build 37.2).

\subsection{Comparative performance}

We summarized paired-end fragments at the feature level, where the features represented the broad regions of all annotated genes.
In this application, a fragment should be counted multiple times if it overlaps multiple genes.

Table~\ref{tab:chipseq} compares the performance of \emph{featureCounts} to that of the \emph{countOverlaps} function of the \emph{IRanges} package, the \emph{htseq-count} script and the \emph{coverageBED} program in the \emph{BEDTools} software suite.
\emph{countOverlaps} was used for this comparison instead of \emph{summarizeOverlaps} because it allows multi-overlap reads to be assigned to multiple features.

\emph{featureCounts} and \emph{countOverlaps} yielded identical counts for every gene, but \emph{featureCounts} was considerably faster and more memory efficient.
\emph{countOverlaps} was also run chromosome by chromosome to save memory.
This reduced the peak memory usage, although it remained more than a hundred times that used by \emph{featureCounts}.
Note that \emph{featureCounts}, unlike \emph{countOverlaps}, can count fragments with only one end successfully mapped, but such fragments were not included in this evaluation to ensure that the timings and memory use for \emph{featureCounts} and \emph{countOverlaps} were for identical operations.

\emph{coverageBED} assigned slightly fewer fragments than \emph{featureCounts}.
We found this was because \emph{coverageBED} used only the first read of each fragment to assign the whole fragment to features.
\emph{htseq-count} counted 7--8\% fewer fragments, presumably because it does not count multi-overlap fragments.
\emph{htseq-count} was run in `intersection-nonempty' mode as well as in `union' mode so as to count more fragments, but this did not make up much of the shortfall.

Columns 3 and 4 of Table~\ref{tab:chipseq} show that \emph{featureCounts} was about five times faster and used about 10 times less memory than the next most efficient tool.

\begin{table}[t]
\caption{Performance results on the H3K27me3 ChIP-seq dataset.
\emph{featureCounts} is the fastest method and uses least memory.
It counts the same number of fragments as \emph{countOverlaps}, but more than \emph{htseq-count} or \emph{coverageBED}.
\label{tab:chipseq}}
{\begin{tabular}{p{5cm}p{3cm}p{3cm}p{3cm}}
\hline
Method & \# fragments & Time (Mins) & Memory (MB) \\
\hline

\emph{featureCounts}	& 5,392,155 & 0.9 & 4 \\

\rule{0pt}{3ex}\emph{countOverlaps}	& 5,392,155 & 24.4 & 7,000 \\
(whole genome at once) \\

\rule{0pt}{2.5ex}\emph{countOverlaps}	& 5,392,155 & 36.6 & 783 \\
(by chromosome) \\

\rule{0pt}{3ex}\emph{htseq-count}		& \multirow{2}{*}{4,978,050}  & \multirow{2}{*}{36.0} & \multirow{2}{*}{31} \\
(union) \\

\rule{0pt}{2.5ex}\emph{htseq-count}		& \multirow{2}{*}{4,993,644}  & \multirow{2}{*}{35.7} & \multirow{2}{*}{31} \\
(intersection-nonempty) \\

\rule{0pt}{3ex}\emph{coverageBED}		& 5,366,902  & 4.4 & 41 \\

\hline\\
\end{tabular}}{
\linebreak
Table shows the total number of fragments counted, time taken and peak memory used.
\emph{featureCounts} was set to count multi-overlap fragments.
Results are shown for \emph{countOverlaps} (i) when run on the whole genome at once and (ii) when run chromosome by chromosome.
Running by chromosome conserves memory but takes longer.
Results are shown for \emph{htseq-count} in two possible counting modes.
For \emph{coverageBED}, the BAM input file was converted to a BED file for summarization using \emph{bamToBed} with options `-bedpe' and `-split'.
}
\end{table}

\section{Performance when the number of reference sequences is large}

\subsection{Simulated data}

Datasets with large numbers of reference sequences are challenging because the read count software must match the contig names of features to those of reads in an efficient manner.
To examine performance under these conditions we simulated reads from an incompletely assembled genome with relatively large number of scaffolds.
We used an assembly of the budgerigar genome generated in the Assemblathon 2 project \citep{assemblathon2,assemblathon2_annot}.
For this assembly there are 16,204 annotated genes with 153,724 exons located on 2,850 scaffolds.
Eight million 100bp single-end reads were randomly extracted from the annotated exonic regions in the assembled scaffolds.
The simulated reads were entered into a SAM file.
Read mapping information was filled according to the locations from where the reads were extracted.

\subsection{Comparative performance}

The simulated reads were then summarized at the gene level.
Table~\ref{tab:sim} compares \emph{featureCounts} to \emph{summarizeOverlaps} and \emph{htseq-count} for this dataset.
As seen before on the RNA-seq data, \emph{summarizeOverlaps} yields the same counts as \emph{featureCounts} while \emph{htseq-count} yields slightly fewer.
\emph{featureCounts} maintained its efficiency advantage over the other methods in this evaluation, increasing its speed advantage over \emph{summarizeOverlaps} in this context.

\begin{table}[t]
\caption{Performance with RNA-seq reads simulated from an annotated assembly of the budgerigar genome.
The annotation includes 16,204 genes located on 2,850 scaffolds.
\emph{featureCounts} is fastest and uses least memory.
\label{tab:sim}}
{\begin{tabular}{p{4.5cm}p{0.1cm}p{2.5cm}p{0.1cm}p{3cm}p{3cm}}
\hline
Methods && \# reads && Time (Mins) & Memory (MB) \\
\hline

\emph{featureCounts}		&& 7,924,065 && 0.6 & 15 \\

\rule{0pt}{3ex}\emph{summarizeOverlaps}	&& \multirow{2}{*}{7,924,065} && \multirow{2}{*}{12.6} & \multirow{2}{*}{2400} \\
(whole genome at once) \\

\rule{0pt}{2.5ex}\emph{summarizeOverlaps}	&& \multirow{2}{*}{7,924,065} && \multirow{2}{*}{53.3} & \multirow{2}{*}{262} \\
(by scaffold) \\

\rule{0pt}{2.5ex}\emph{htseq-count}			&& 7,912,439 && 12.1 & 78 \\

\hline\\
\end{tabular}}{
\linebreak
Table gives the total number of reads counted, time taken and peak memory used. \emph{htseq-count} was run in `union' mode.}
\end{table}

\section{Theoretical analysis of algorithmic complexity}

This section gives a theoretical analysis of the computational time and memory storage required by \emph{featureCounts} and the other algorithms.
The actual time and memory consumed by a computer program depends on the computer hardware, operating system and other factors as well as on the mathematical efficiency of the algorithm used.
However we can derive theoretical expressions for the rate at which time and memory used by any specific algorithm should increase with the number of reads, the number of features and the density of features in the genome.
The time complexity of the \emph{featureCounts} algorithm can be derived as $O(f\log{f}+r\sqrt{k_1})$, where $f$ is the number of features, $r$ is the number of reads and $k_1$ is the number of features included in a genomic bin.
This means that the number of elementary computations used by the algorithm increases linearly with the number of reads, independently of the number of features, and somewhat faster than linearly with the number of features.
The space complexity of the \emph{featureCounts} algorithm is $O(f+b_1)$, meaning that memory used increases linearly with the number of features plus the number of bins $b_1$.
Time and space complexities for all the algorithms are given in Table~\ref{tab:complexity}.

The number of reads is typically large, so rate of increase with $r$ is especially important.
The \emph{featureCounts} algorithm has the lowest time complexity of the algorithms being compared.
The red-black tree search algorithm used by \emph{htsesq-count} has higher complexity because $\log f$ is typically larger than the square-root of the number of features per bin used by \emph{featureCounts}.
The hierarchical search within bins used by \emph{featureCounts} is more efficient than the sequential search carried out by \emph{coverageBED}, because most reads overlap multiple levels of bins with \emph{coverageBED} causing $k_2$ to be typically greater than $k_1$.
\emph{countOverlaps} and \emph{summarizeOverlaps} sort reads according to their mapped locations and then use an interval tree to find features overlapping with reads.
The sort step is especially expensive and introduces $r\log r$ terms.

The \emph{htseq-count} algorithm has the best theoretical space complexity, but \emph{featureCounts} is not far behind because the number of bins $b_1$ is usually small compared to $f$.
\emph{BEDTools} has a higher space complexity than \emph{featureCounts} because it uses more bins.
\emph{CountOverlaps} and \emph{summarizeOverlaps} have higher space complexities that depend on the number of reads as well as on the number of features.

In practice, the running time and memory usage of a software program are determined not just by the inherent time and space complexities of the algorithm it adopts but also by the efficiency of the software implementation.
The practical timings show that \emph{featureCounts} achieves further efficiency gains from high performance C programming and direct memory manipulation.

\begin{table}
\caption{Theoretical time and space complexity.
The table gives proportionality factors for the number of computations (time complexity) and memory locations (space complexity) required by each algorithm.
Time complexities depend on the number of features $f$, the number of reads $r$ and the number of features included in genomic bins overlapping the query read, $k$.
Space complexity also depends on the number of bins, $b$.
\label{tab:complexity}}
{\begin{tabular}{p{4.5cm}p{6cm}p{4cm}}
\hline
Method & Time & Memory \\
\hline
\emph{featureCounts}			& $f\log{f}+r\sqrt{k_1}$ & $f+b_1$ \\
\emph{countOverlaps} & $f\log{f}+r\log{r}+r$ & $f+r$\\
\emph{summarizeOverlaps} & $2f\log{f}+2r\log{r}+r+f$ & $f+r$\\
\emph{htseq-count}       & $f\log{f}+r\log{f}$ & $f$\\
\emph{coverageBED}       & $f\log{f}+rk_2$ & $f+b_2$\\
\hline\\
\end{tabular}}{
\linebreak
Complexities are interpretted as $O(x)$ where $x$ is the expression given in the table.
The number of bins used by \emph{coverageBED}, $b_2$, is greater than the number of bins used by \emph{featureCounts}, $b_1$.
The number of within-bin features $k_2$ for \emph{coverageBED} is typically greater than $k_1$ for \emph{featureCounts}.}
\end{table}

\section{Discussion}

Read summarization is an important step in many next-gen sequencing data analyses.
In this study, we developed a new read summarization program called \emph{featureCounts} and compared it with existing methods in terms of efficiency and accuracy.
Our results showed a high concordance between alternative methods in summarization accuracy.
However, there was a large difference observed in their computational cost.
The \emph{featureCounts} method was found to be an order of magnitude faster on average and far more memory-efficient than other methods.
The very high computational efficiency of \emph{featureCounts} is due to its ultrafast feature search algorithm and its highly efficient implementation entirely using the C programming language.

All results presented in this article were produced using a single thread, but \emph{featuresCounts} also supports multithreaded processing, making it particularly useful for summarizing data generated in large sequencing studies.
It is the only existing read count method that supports multithreading.

This program provides a wide range of options to allow users to fully control how their read data can be best summarized.
Users can choose whether or not they should count the reads that overlap with more than one feature or meta-feature.
This choice is often determined by the experiment type.
Reads overlapping with more than one gene (a meta-feature) should not be counted in a RNA-seq experiment because such reads can only originate from one gene, but usually they should be counted in a gDNA-seq experiment such as a histone ChIP-seq experiment.
This program also allows users to filter out reads before summarization using a number of metrics such as mapping quality scores, fragment mappability (whether two ends from the same fragment are both successfully mapped or not), fragment length, strandness, chimerism and so on.
It can automatically detect either SAM or BAM format read input.
It also allows users to specify whether those reads which were reported with more than mapping location (multi-mapping) should be counted or not.
Many of these useful features are not supported by other programs.

The \emph{featureCounts} program has been implemented in both SourceForge \emph{Subread} package \citep{subread_sf} and Bioconductor \emph{Rsubread} package \citep{Rsubread}.
The R function provides users with an R interface so that they can access this program from their familiar R environment.
It calls the underlying compiled C program to perform all the read summarization operations, and hence
has the same speed and memory usage as that of the SourceForge \emph{Subread} package which is written entirely in C.
The implementation of \emph{featureCounts} in R enables complete pipelines to be established for analysing next-gen sequencing data using Bioconductor software programs.
For example, functions included in Bioconductor packages \emph{Rsubread}, \emph{limma} and \emph{edgeR} can be used to perform complete RNA-seq and histone ChIP-seq analyses, starting from read mapping, to read summarization and finally to differential expression analyses or differential histone modification analyses.
Due to its high efficiency and accuracy, we believe the \emph{featureCounts} program will be a useful tool in the bioinformatics toolbox for analysing next-gen sequencing data.

\section{ACKNOWLEDGEMENTS}

We thank Leming Shi and Charles Wang for providing the SEQC pilot data, and Aaron Lun for helpful comments.

\section{FUNDING}

This work was supported by a Project Grant (1023454) and a Fellowship (to GKS) from the Australian National Health and Medical Research Council (NHMRC).
It was made possible through Victorian State Government Operational Infrastructure Support and Australian Government NHMRC IRIIS.

\bibliographystyle{natbib}
\bibliography{featureCounts}

\end{document}